\documentclass[aps,showpacs,preprintnumbers,amsmath,amssymb,a4paper,preprint]{revtex4}
\usepackage{graphicx}
\usepackage{dcolumn}
\usepackage{bm}
\usepackage{color}

\begin{document}

\title{Photoelectron circular dichroism in the multiphoton ionization by short laser pulses:
III. Photoionization of fenchone in different regimes}

\author{\firstname{Anne~D.} \surname{M\"{u}ller}}
\affiliation{Institute of Physics and CINSaT, University of Kassel, Heinrich-Plett-Str.~40, 34132 Kassel, Germany}

\author{\firstname{Eric} \surname{Kutscher}}
\affiliation{Institute of Physics and CINSaT, University of Kassel, Heinrich-Plett-Str.~40, 34132 Kassel, Germany}

\author{\firstname{Anton~N.} \surname{Artemyev}}
\affiliation{Institute of Physics and CINSaT, University of Kassel, Heinrich-Plett-Str.~40, 34132 Kassel, Germany}

\author{\firstname{Philipp~V.} \surname{Demekhin}}\email{demekhin@physik.uni-kassel.de}
\affiliation{Institute of Physics and CINSaT, University of Kassel, Heinrich-Plett-Str.~40, 34132 Kassel, Germany}

\date{\today}

\begin{abstract}
Photoelectron circular dichroism (PECD) in different regimes of multiphoton ionization of fenchone is studied theoretically using the time-dependent single center method. In particular, we investigate the chiral response to the one-color multiphoton or strong-field ionization by circularly polarized 400 and 814~nm optical laser pulses or 1850~nm infrared pulse. In addition, the broadband ionization by short coherent circularly polarized 413--1240~nm spanning  pulse is considered. Finally, the two-color ionization by the phase-locked 400 and 800~nm pulses, which are linearly polarized in mutually-orthogonal directions, is investigated. The present computational results on the one-color multiphoton ionization of fenchone are in agreement with the available experimental data. For the ionization of fenchone by broadband and bichromatic pulses, the present theoretical study predicts substantial multiphoton PECDs.
\end{abstract}


\maketitle

\section{Introduction}
\label{sec:Intro}

Since its first observation in the pioneering experiments with femtosecond laser pulses \cite{Lux12AngChm}, different aspects of the multiphoton photoelectron circular dichroism (PECD) were studied in many independent experiments with different chiral molecules \cite{Lehmann13jcp,Ram13EPJ,Lux15CPC,Janssen14,Nanosec19,Lux16ATI,Beaulieu16NJP, Beaulieu16td,Cireasa15,Comby16td,Beaulieu17as,Beaulieu18PXCD, Rafiee16wl,Kastner17wl,Comby16elee,Kastner16ee}. At present, it is understood that multiphoton PECD is a universal chiroptical effect which persists in many ionization regimes:  resonance-enhanced multiphoton ionization \cite{Lux12AngChm,Lehmann13jcp,Ram13EPJ,Lux15CPC,Janssen14,Nanosec19}, above-threshold  \cite{Lux16ATI,Beaulieu16NJP,Beaulieu16td} and strong-field tunnel \cite{Cireasa15,Beaulieu16NJP,Beaulieu16td} ionizations, as well as in multiphoton ionization by different two-color pump-probe schemes \cite{Beaulieu16td,Comby16td,Beaulieu17as,Beaulieu18PXCD}. The substantial size of this chiroptical asymmetry induced at different laser pulse wavelengths \cite{Nanosec19,Beaulieu16NJP,Beaulieu16td,Rafiee16wl,Kastner17wl} and its intricate dependence on the ellipticity of light \cite{Lux15CPC,Comby16elee} makes multiphoton PECD a powerful laboratory tool for the real time \cite{Comby16elee}  enantiomeric excess determination with up to a sub-percent accuracy \cite{Kastner16ee}.

Theoretical investigation of the multiphoton PECD is a very challenging task from both, conceptual and computational aspects \cite{Lehmann13jcp,Cireasa15,Beaulieu18PXCD,Dreissigacker14,Goetz17,Goetz19,Rozen19,TDSC1,TDSC2}. In our previous works Refs.~\cite{TDSC1} and \cite{TDSC2} (hereafter referred to as Papers I and II, respectively), a nonperturbative theoretical method to study angle-resolved multiphoton ionization of polyatomic molecules was developed. This Time-Dependent Single Center (TDSC) method consists in the propagation of single-active-electron wave packets driven by laser pulses in the molecular potential, and it solves the time-dependent Schr\"{o}dinger equation by the efficient numerical approach of Refs.~\cite{Demekhin13H,Artemyev16He1,Artemyev17He2,Artemyev17He3,Artemyev18He4}. Details on the numerical implementation of the TDSC method can be found in Paper~I~\cite{TDSC1}, where it was tested on the one-photon ionization and two-photon above-threshold ionization of a model methane-like chiral system. In Paper~II~\cite{TDSC2}, the TDSC method was successfully applied to study three-photon resonance-enhanced ionization and four-photon above-threshold ionization of fenchone and camphor molecules by circularly polarized laser pulses.

In the present work, we apply this theoretical method to study ionization of fenchone in different multiphoton regimes. In Sections~\ref{sec:multiphoton} and \ref{sec:tunnel}, we follow a passage from the multiphoton ionization  by optical laser pulses to the strong-field ionization  by intense infrared pulses and compare our theoretical results with recent experimental results from Refs.~\cite{Beaulieu16NJP,Beaulieu16td}. In Section~\ref{sec:broadband}, we investigate a possibility to induce a multiphoton PECD by short coherent broadband laser pulses, which evoke interference between the continuous manifold of multiphoton ionization pathways available at once. Finally in Section~\ref{sec:bichromatic}, we study PECD induced by six-red-photon vs three-blue-photon ionization with Lissajous-type bichromatic fields, which is predicted in the recent theoretical work~\cite{PRLw2w}.

\section{Results and discussion}
\label{sec:results}

The computational details on the application of the TDSC method to fenchone can be found in Section~II of Paper~II~\cite{TDSC2}. Therefore, only essential points relevant to the present study, as well as details which are specific to each ionization regime will be outlined below. All calculations were performed in the dipole-velocity gauge, which enforces convergence of the numerical solution over the partial waves \cite{VG1,VG2}. In this gauge and for the presently considered peak intensities below $4\times 10^{13}$~W/cm$^2$, partial wave expansions of the photoelectron wave packets with $\ell,\vert m\vert \leq 25$, which were used in Paper~II~\cite{TDSC2}, are sufficient to ensure convergence of the numerical solution over the angular momentum quantum numbers \cite{VG2}. { The radial grid was divided into finite elements of different size in the inner and outer regions \cite{TDSC2}, each covered by normalized Lagrange interpolating polynomials constructed over 10 Gauss-Lobatto points \cite{Demekhin13H,Artemyev16He1,Artemyev17He2,Artemyev17He3,Artemyev18He4}. The time-dependent electron wave packets were propagated using the short-iterative Lanczos method \cite{PropL}. Electrons with very high kinetic energies were absorbed by a mask function at the end of the radial grid \cite{Artemyev17He2,Artemyev17He3}.} In order to obtain photoelectron momentum distributions, here the final wave packets were projected on Coulomb waves \cite{Cont1,Cont2,Cont3,Cont4,Cont5}. This yields more accurate electron spectra than plane waves used in Paper~II~\cite{TDSC2}.

\subsection{Multiphoton ionization}
\label{sec:multiphoton}

We first discuss the photoionization (PI) and the above-threshold ionization (ATI) of fenchone with 400~nm laser pulses. For computational reasons, we employed   sine-squared pulses $g(t)=\sin^2\left(\frac{\pi t}{T}\right)$, which have well-defined beginning, end, and full duration $T$. Here, we used the same total pulse duration $T=20$~fs (supports about 15 optical cycles), peak intensity $10^{12}$~W/cm$^2$, and radial grid of $r \leq 425$~a.u., as in Paper~II~\cite{TDSC2}. The molecular frame $z^\prime$-axis of fenchone was chosen along its C=O bond, and propagation was performed at different orientation Euler angles $\alpha$ and $\beta$ in steps of $\Delta \alpha= \Delta \beta=0.1\,\pi$. Owing to the axial symmetry of the circularly polarized field and absence of the carrier-envelope phase (CEP) asymmetry, the third Euler angle $\gamma$ is irrelevant.

\begin{figure}
\includegraphics[scale=0.6]{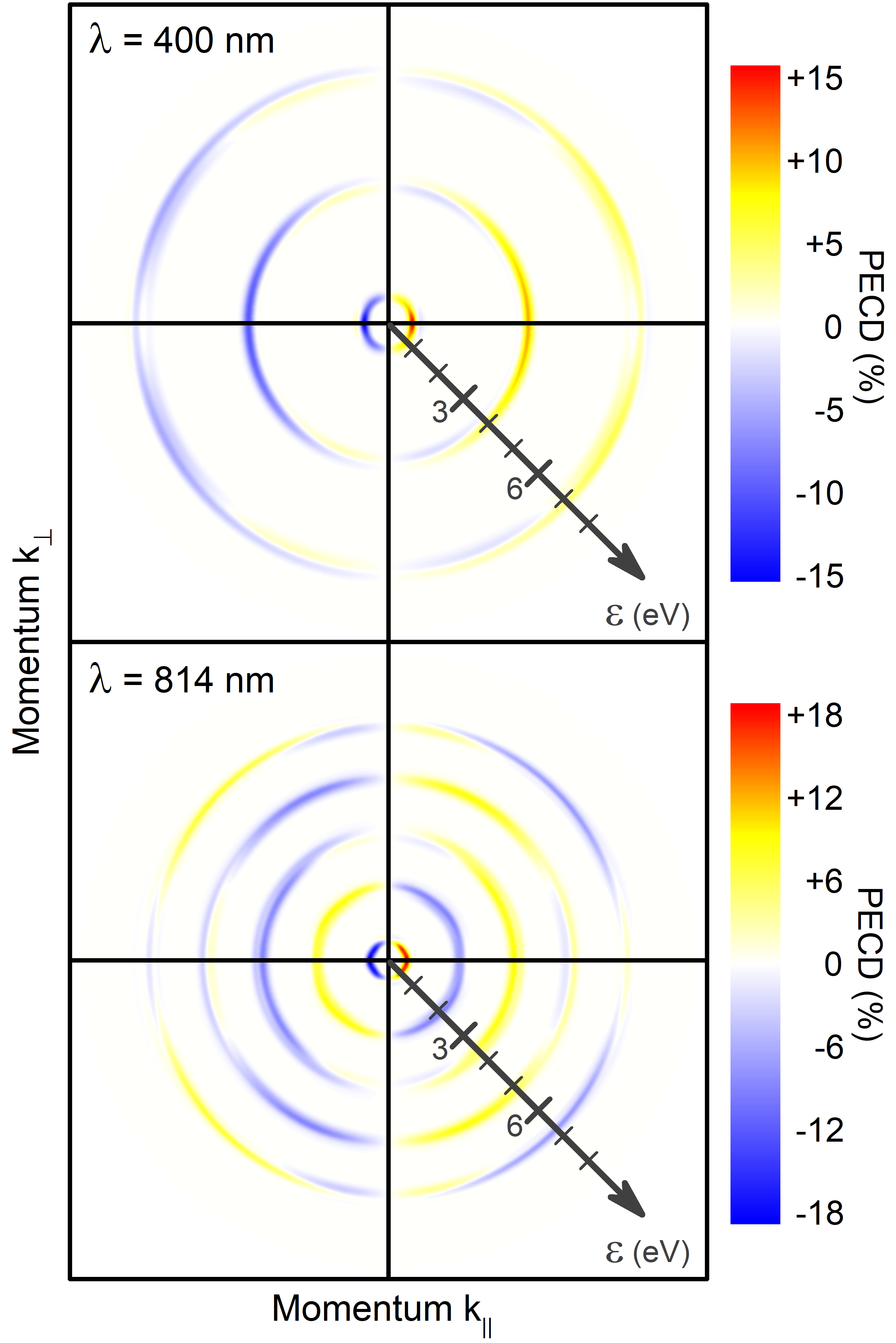}
\caption{Multiphoton PECD, computed in the present work for randomly-oriented R(--) fenchone molecules exposed to 400~nm (upper panel) and 814~nm (lower panel) laser pulses. The three rings in the upper panel represent photoelectrons released by the three-photon ionization and the four- and five-photon ATI processes. The five rings in the lower panel represent photoelectrons released by the six-photon ionization and seven- to ten-photon ATI processes.  Shown is the relative difference between the two electron spectra computed for the right-handed  and for the left-handed  circular polarizations. Each PECD signal is shown in percent relatively to the maximal pixel intensity in the respective PI or ATI spectrum computed for one polarization. The pulse propagates horizontally from the left to the right, i.e., the laboratory $z$-axis coincides with $k_{||}$. Note that the photoelectron momentum $k$ is set on the linear scale of the kinetic energy $\varepsilon$.}\label{fig:fenMulti}
\end{figure}

Upper panel of Fig.~\ref{fig:fenMulti}  depicts the presently computed multiphoton PECD in the momentum space ($k_{||}$,$k_\bot$). Shown is the normalized (see below) difference between the two emission distributions, computed for the right- and left-handed  circularly polarized  pulses. In this plot, the photoelectron emission angle  $\theta$ is defined with respect to $k_{||}$, and the forward/backward emission directions correspond to the right/left sides of the figure. For a more convenient representation of the data, the photoelectron momentum is shown in the energy scale in electron volts (see the downward inclined arrow). As follows from the energy conservation $\varepsilon=k^2/2=n\omega-IP$ with $\omega=3.10$~eV and $IP=8.6$~eV \cite{Powis08}, the three distinct photoelectron signals at $\varepsilon=0.70$, 3.8, and 6.9~eV correspond to the three-photon PI (3PI), and four- and five-photon ATI (4ATI and 5ATI) of fenchone, respectively. Since the intensities of these three signals differ by about two orders of magnitude, the respective PECDs are shown in percent of the maximal intensity of each spectrum.

As one can see from the upper panel of Fig.~\ref{fig:fenMulti}, the 3PI and also 4ATI and 5ATI electrons, released from R(--) fenchone molecules by the 400~nm circularly polarized pulses, altogether, exhibit a positive PECD. This fact is in agreement with the experimental results of Refs.~\cite{Beaulieu16NJP,Beaulieu16td} (see Fig.~2a in Ref.~\cite{Beaulieu16NJP} and Fig.~6f in Ref.~\cite{Beaulieu16td}; note also that the measured PECD of S(+) fenchone has an opposite sign). Even sizes of the presently computed  chiral asymmetries, of about 13\%, 9\%, and 5\% in the maxima of 3PI, 4ATI, and 5ATI signals, respectively, are in good agreement with those observations {(note that the maximal values of the computed PECD are slightly different from the values of the integral effect \cite{TDSC1,TDSC2}, $\widetilde{PECD}=2b_1 -\frac{1}{2}b_3 +\frac{1}{4}b_5 -\frac{5}{32}b_7 + ... $, which are equal to about +10\%, +4.0\% and +3.1\%, respectively).} Indeed, Fig.~2c of Ref.~\cite{Beaulieu16NJP} reports a similar decrease of the total PECD for HOMO electron from about 12\% for the 3PI to about 7\% and further to 3\% for the 4ATI and 5ATI (shown by lightest shaded area there). Finally, the presently computed PECDs for 3PI and  4ATI signals are in a very good agreement with the respective results of the independent experiments performed by different authors \cite{Lux12AngChm,Lux15CPC,Lux16ATI}.

Following the logic of Refs.~\cite{Beaulieu16NJP,Beaulieu16td}, we studied PI and ATI of fenchone by circularly polarized 800~nm laser pulses. For the present atomic-like electron energy spectrum of fenchone, the carrier frequency of this pulse, which is resonant for the four-photon transition  (as that of 400~nm pulse is resonant for the two-photon transition, see Paper~II~\cite{TDSC2}), was accidentally resonant for the absorption of a subsequent fifth photon. Therefore, even at rather moderate peak intensity of $10^{12}$~W/cm$^2$, this one-photon resonant transition in the fifth photon absorption step turned into the Rabi-floppings and yielded substantial Autler-Towns energy splittings of the PI and all ATI peaks. One way to avoid the undesired Autler-Towns doublets is to considerably reduce the peak intensity of the 800~nm pulse. Here, however, we decided to use a slightly detuned wavelength of 814~nm and to keep the pulse intensity as high as sufficient to clearly observe higher ATI peaks. To further avoid any CEP asymmetry, we used somewhat longer pulses with the full  duration of $T=30$~fs (support about 10 optical cycles). Finally, we enlarged the radial grid to $r \leq 600$~a.u. to enable support of fast ATI electron released by up to ten-photon absorption during the whole propagation time $T$.

Results of the calculations performed for the 814~nm pulses are depicted in the lower panel of Fig.~\ref{fig:fenMulti} in the same way as for the 400~nm pulses. Five different PECD signals, seen in this panel at the electron kinetic energies of about $\varepsilon=0.5$, 2.0, 3.6, 5.1 and 6.6~eV, correspond to the six-photon PI (6PI) and seven- to ten-photon ATI processes (7ATI -- 10ATI), respectively. The PECD for  6PI is positive and reaches about 18\% of the maximal PI intensity. For the 7ATI, the PECD is negative and reaches up to about 10\% in its maximum. Further on, PECDs of the 8ATI and 9ATI signals are positive reaching about 9\% and 8\% at their maxima, respectively. The 10ATI signal is almost three orders of magnitude weaker than that of 6PI, and it exhibits a negative PECD of up to about 7\% of its own intensity. {
The integral $\widetilde{PECDs}$ are equal to $+18\%$, $-14\%$, $+8.6\%$, $+13\%$ and $-4.5\%$ for the 6PI and 7ATI -- 10ATI, respectively. The fact that different ATI spectra exhibit different sign of the chiral asymmetry suggests that the respective electrons experience considerable multiple scattering effects.}

The present theoretical results obtained for the 814~nm pulse are in a reasonable agreement with the experimental results  of Refs.~\cite{Beaulieu16NJP,Beaulieu16td} acquired for the 800~nm pulse (see Fig.~3d in Ref.~\cite{Beaulieu16NJP} and Fig.~6h in Ref.~\cite{Beaulieu16td}). All experimentally observed 7ATI -- 10ATI signals exhibit a substantial PECDs reaching about 8\%~\cite{Beaulieu16NJP} (note that the 6PI peak is not present in the experimental spectrum). To guide the eye, the energy positions of the ATI signals, resulting from the ionization of HOMO electron, are marked in Fig.~6h of Ref.~\cite{Beaulieu16td} by solid circles. The 7ATI peak exhibits the strongest PECD of the same sign as in the present theory (note that Fig.~6h of Ref.~\cite{Beaulieu16td} shows data for S(+) fenchone). Similar to the present theory (lower panel of  Fig.~\ref{fig:fenMulti}), PECDs of the 8ATI and 9ATI peaks change signs as compared to that of the 7ATI signal.

For the very weak 10ATI signal, the theoretical and experimental PECDs exhibit opposite signs. This disagreement could be related to the fact that an almost one order of magnitude stronger pulse intensity of $9\times 10^{12}$~W/cm$^2$ was used in the experiments of Refs.~\cite{Beaulieu16NJP,Beaulieu16td}. Indeed, the 10ATI spectrum can be clearly seen in Fig.~6c of Ref.~\cite{Beaulieu16td} at $\theta=90^\circ$, while in the present calculations utilizing $10^{12}$~W/cm$^2$ pulses it is almost three orders of magnitude weaker than the 6PI spectrum). In addition, stronger pulses used in Refs.~\cite{Beaulieu16NJP,Beaulieu16td} cause a slight blurring of the spectrum and an additional angular structuring of the PECD signals of all ATI peaks. Such an angular structuring of PECD appears only in the presently computed 9ATI and 10ATI signals.

\subsection{Strong-field ionization}
\label{sec:tunnel}

\begin{figure}
\includegraphics[scale=0.6]{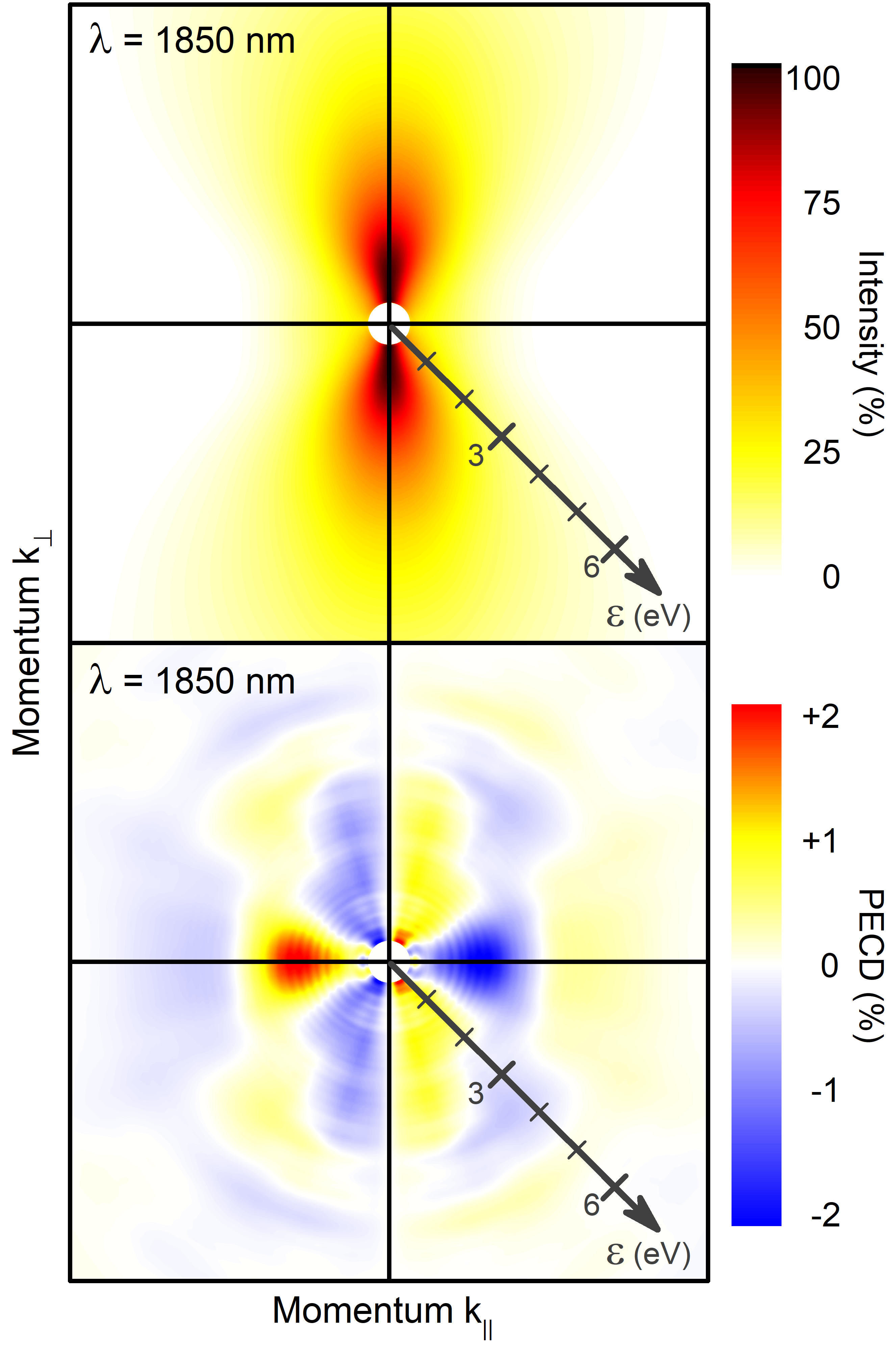}
\caption{\emph{Upper panel:} The spectrum of electrons computed for the strong-field ionization of randomly-oriented R(--) fenchone molecules by the right-handed circularly polarized, 1850~nm,  sine-squared laser pulse of 50~fs full duration and $4\times 10^{13}$~W/cm$^2$ peak intensity. The corresponding Keldysh parameter of $0.58$ indicates that the ionization takes place in the intermediate regime between the tunnel  and multiphoton ionization extremes. The maximal pixel intensity in spectrum is set to 100\%. Note that for chosen propagation times, the slow photoelectrons with kinetic energies below 0.4~eV cannot be separated from the highly-excited bound electronic states. \emph{Lower panel:}  The respective multiphoton PECD in percent of the maximal pixel intensity (see caption of Fig.~\ref{fig:fenMulti} for details on the data presentation). }\label{fig:fenTunnel}
\end{figure}

We now turn to the strong-field ionization of fenchone by circularly polarized 1850~nm infrared pulses. In the calculations, we used the same peak intensity of $4\times 10^{13}$~W/cm$^2$ as in the experiments of Refs.~\cite{Beaulieu16NJP,Beaulieu16td}. Thereby, we consider an intermediate regime between the multiphoton and tunnel ionization, which is characterized by the Keldysh parameter of $0.58$. Owing to  computational reasons, the full duration of the sine-squared pulses was set to $T=50$~fs. Such  pulses support more than eight optical cycles, which causes a very small CEP asymmetry of the field. In order to support photoelectrons with the kinetic energies of up to about 15~eV during the whole propagation time, we used the radial grid of $r \leq 1200$~a.u. To further reduce computational efforts, we covered the molecular orientation intervals of $\alpha\in[0,2\pi)$ and $\beta\in[0,\pi]$ by larger steps of $\Delta \alpha= \Delta \beta= 0.2\,\pi$. The results of the present calculations are collected in Fig.~\ref{fig:fenTunnel}.

The spectrum of photoelectrons released by the strong 1850~nm pulses is depicted in the upper panel of Fig.~\ref{fig:fenTunnel} in the momentum space, while the respective PECD is shown in the lower panel in the similar way as for the multiphoton ionization regime in Fig.~\ref{fig:fenMulti}. The photoelectron spectrum in the upper panel of this figure is typical for the tunnel ionization regime: It is aligned vertically since photoelectrons tunnel preferably along the electric field which rotates in the dipole plane. The present spectrum in the upper panel of Fig.~\ref{fig:fenTunnel} is very similar to that obtained in the experiments of Refs.~\cite{Beaulieu16NJP,Beaulieu16td} (see Fig.~3c in Ref.~\cite{Beaulieu16NJP} and Fig.~6e in Ref.~\cite{Beaulieu16td}). We note, that after the full propagation time, photoelectron wave packets with low kinetic energies below 0.4~eV still overlap with highly-excited bound electronic states of the molecule and cannot be analyzed in a reliable way. Therefore, this part of the spectrum is excluded from Fig.~\ref{fig:fenTunnel}.

As can be seen from the  lower panel of Fig.~\ref{fig:fenTunnel}, the strong 1850~nm circularly polarized pulses induce a noticeable PECD (on the order of 2\%). A chiral asymmetry of similar size of around 2\% is also observed in Refs.~\cite{Beaulieu16NJP,Beaulieu16td}. {In Ref.~\cite{Beaulieu16NJP}, it is qualitatively reproduced by means of classical trajectory calculations.} Both, the presently computed and the measured  PECD signals (see Fig.~3f in Ref.~\cite{Beaulieu16NJP} and Fig.~6j in Ref.~\cite{Beaulieu16td}) are aligned along the dipole plane. The measured PECD exhibits one sign for all photoelectron kinetic energies, while the presently computed PECD changes its sign twice. Interestingly, the inner part of the presently computed PECD signal, which corresponds to the most intense part of the spectrum, qualitatively reproduces the experimental results, including the sign of the measured asymmetry (note that Fig.~3f in Ref.~\cite{Beaulieu16NJP} and Fig.~6j in Ref.~\cite{Beaulieu16td} show data for S(+) fenchone). The computed PECD changes its sign  for larger photoelectron energies in the low-intensity part of the spectrum.

\subsection{Broadband ionization}
\label{sec:broadband}

So far, we discussed PECD induced through a particular multiphoton ionization pathway governed by the laser pulses, which carry photons in a narrow energy interval  around their carrier frequencies.  What happens if a chiral molecule is exposed to circularly polarized coherent laser pulses which support photons in a broad energy range? In this case, an infinite number of different multiphoton ionization pathways will be available at once, and the respective transition amplitudes will superimpose and interfere. Such kind of interference between different multiphoton pathways is at the heart of coherent control \cite{Goetz19,CoCo1,CoCo2}. In this section, we report theoretical predictions of a multiphoton PECD of fenchone induced by short coherent broadband laser pulses.

In the present study, we used circularly polarized pulses with constant carrier frequencies $\omega_0$ and global phases $\phi_0$, such that all photons supported by the pulses are available at the same time (i.e., without ordering of frequencies in time). Furthermore, the pulses have a symmetric (see also below) energy spectrum and support photons in the well-defined energy interval $\omega\in[\omega_1,\omega_2]$ with $\omega_0=\frac{1}{2}(\omega_2+\omega_1)$. In particular, we chose the sine-squared intensity spectrum of the pulses:
\begin{equation}
I(\omega)=\sin^2\left(\pi\frac{\omega-\omega_1}{\omega_2-\omega_1}\right), ~~\omega\in[\omega_1,\omega_2].
\label{eq:enspectr}
\end{equation}
This energy spectrum is smooth, and it excludes UV photons with the energies above $\omega_2$ and IR photons with the energies below $\omega_1$. In the calculations, we set $\omega_1=1.0$~eV, $\omega_2=3.0$~eV, and $\omega_0=2.0$~eV (corresponds to the 413--1240~nm wavelengths range and $\lambda_0=620$~nm).

\begin{figure}
\includegraphics[scale=.43]{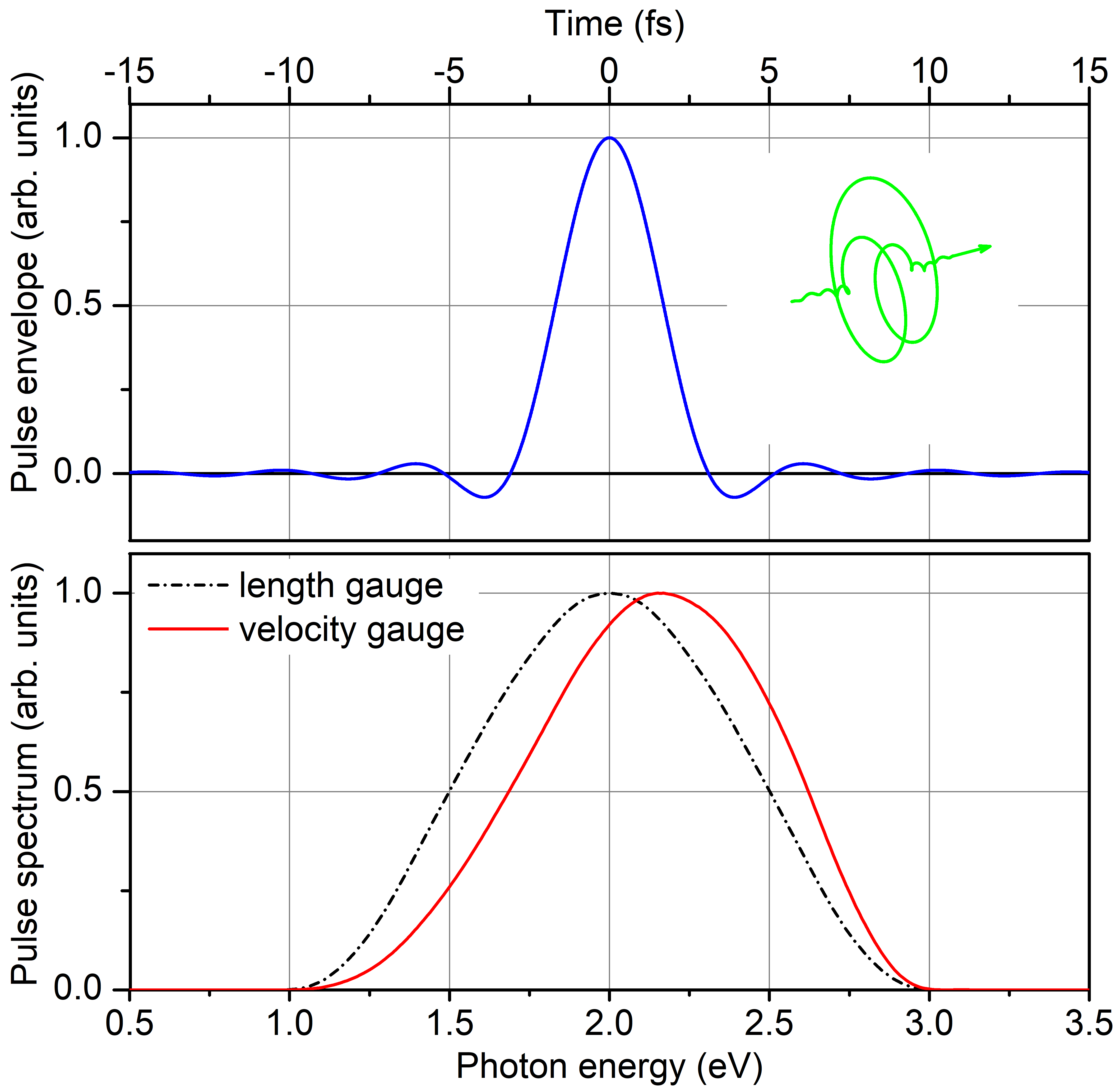}
\caption{\emph{Upper panel:} Time-envelope of the broadband laser pulse defined via Eq.~(\ref{eq:timesp}) with $\omega_1=1.0$~eV and $\omega_2=3.0$~eV. The shape of such few-cycle circularly polarized pulse with the carrier frequency of $\omega_0=2.0$~eV is demonstrated in the inset to this panel. \emph{Lower panel:} Energy spectrum of the broadband laser pulse. In the length gauge (black dash-dotted curve), when the time-envelope defines  electric field vector $\mathcal{E}(t)$, the pulse spectrum is given by Eq.~(\ref{eq:enspectr}) and is symmetric with respect to the chosen carrier frequency $\omega_0=2.0$~eV.  In the velocity gauge (red solid curve), when the time-envelope defines the vector potential $\mathcal{A}(t)$, the spectrum of Eq.~(\ref{eq:enspectr}) is additionally multiplied with the $\omega^2$-function.}\label{fig:fenBBpulse}
\end{figure}

The energy spectrum (\ref{eq:enspectr}) is depicted in the lower panel of Fig.~\ref{fig:fenBBpulse} by the black dash-dotted curve. The time-envelope $g(t)$ of the electric field vector $\mathcal{E}(t)$ of this pulse can be found by the inverse Fourier transformation of the electric field $\mathcal{E}(\omega)=\sqrt{I(\omega)}$ of Eq.~(\ref{eq:enspectr}). For the symmetric energy spectrum (\ref{eq:enspectr}), the respective time-envelope $g(t)$ is also symmetric and real, and it has the following analytic expression:
\begin{equation}
g(t)=\frac{\pi^2}{\pi^2-(\omega_2-\omega_1)^2 t^2}\cos\left(\frac{\omega_2-\omega_1}{2}t\right).
\label{eq:timesp}
\end{equation}
For the presently used parameters, the time-envelope (\ref{eq:timesp}) is depicted in the upper panel of Fig.~\ref{fig:fenBBpulse}. As one can see, the main pulse intensity is concentrated in a narrow interval of a few femtoseconds with long and weak tails on its rising and falling edges. For the carrier frequency of $\omega_0=2.0$~eV, the main part of this pulse supports almost three optical cycles (see inset in the upper panel of Fig.~\ref{fig:fenBBpulse}).

It should be reminded that the present calculations were performed in the dipole-velocity gauge which requires time-evolution of the vector potential $\mathcal{A}(t)$ with $\mathcal{E}(t)=-\partial_t\mathcal{A}(t)$. Reconstruction of the vector potential which yields the energy spectrum (\ref{eq:enspectr}) is not a straightforward analytical task. Therefore, for simplicity, we used the time-envelope (\ref{eq:timesp}) to define time-dependence of the vector potential of circularly polarized broadband pulses. The resulting pulse spectrum, which can be found as $\omega^2I(\omega)$ with $I(\omega)$ from Eq.~(\ref{eq:enspectr}), is depicted in the lower panel of  Fig.~\ref{fig:fenBBpulse} by the red solid curve. As one can see, this energy spectrum is slightly asymmetric with the maximum at $\omega=2.15$~eV.

As can be seen from the inset in the upper panel of  Fig.~\ref{fig:fenBBpulse}, CEP plays a role for such few-cycle circularly polarized pulses. Indeed, the phase $\phi_0$ defines the azimuthal angle $\varphi$ at which the electric field $\mathcal{E}(t)$ has its maximal value. As a consequence, the respective angular emission distribution is not fully axially symmetric. In order to simulate a gas of freely-rotating molecules, the electron spectra computed for a given phase $\phi_0$ need to be averaged over all three orientation Euler angles ($\alpha,\beta,\gamma)$. To reduce computational efforts, we covered orientation intervals of $\alpha\in[0,2\pi)$ and $\beta\in[0,\pi]$ in steps of $\Delta \alpha= \Delta \beta= 0.2\,\pi$. The third orientation angle $\gamma$  defines rotation of the molecular $z^\prime$-axis around the laboratory $z$-axis. For a rather small axial asymmetry of the present few-cycle pulses, it was sufficient to cover the interval of $[0,2\pi)$ by larger steps of  $\Delta \gamma= 0.5\,\pi$.

\begin{figure}
\includegraphics[scale=0.6]{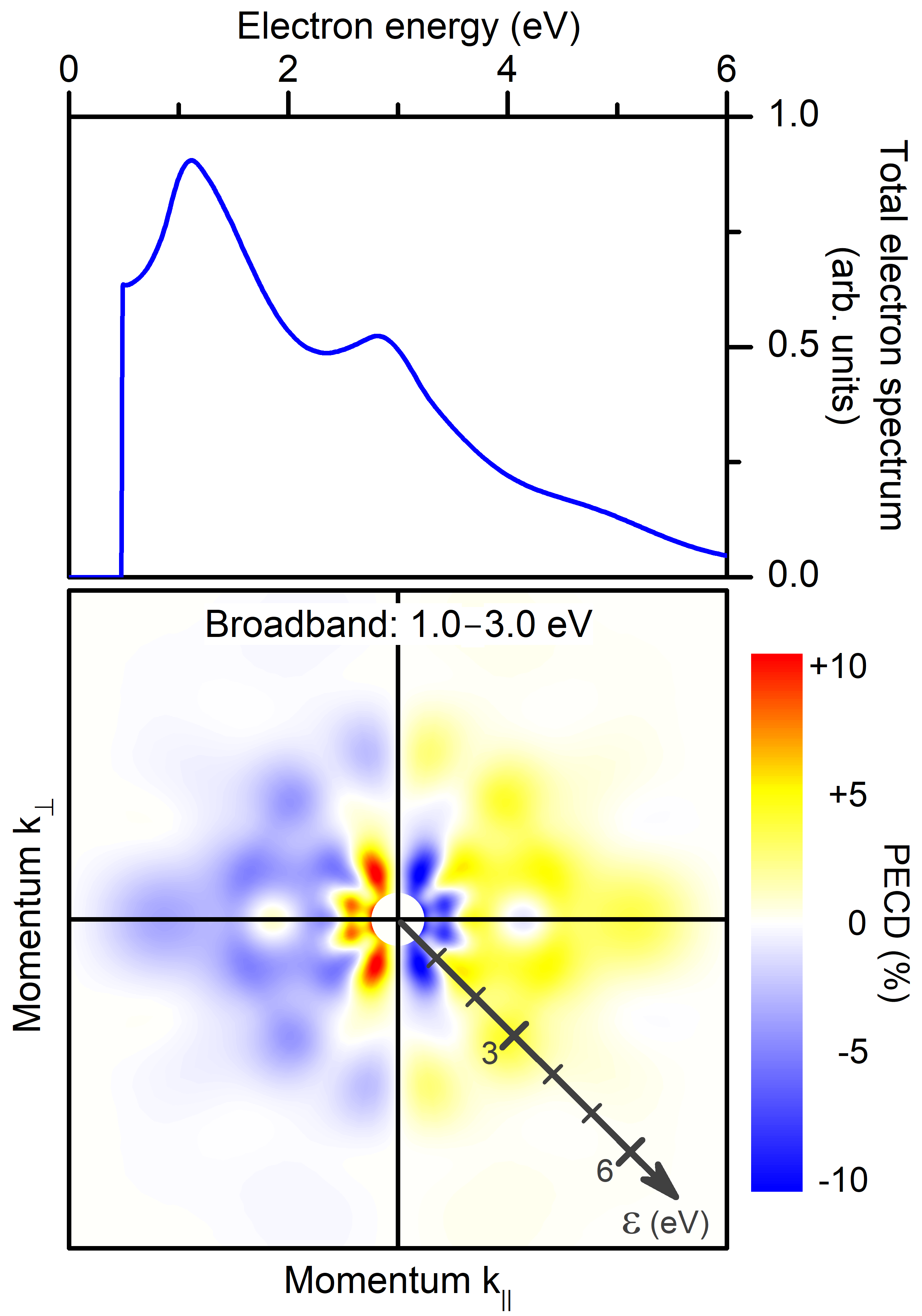}
\caption{\emph{Upper panel:} The total spectrum of electrons computed for randomly-oriented R(--) fenchone molecules exposed to circularly polarized broadband pulses with the time-envelope of the vector potential $\mathcal{A}(t)$ defined via Eq.~(\ref{eq:timesp}), $\omega_1=1.0$~eV and $\omega_2=3.0$~eV, and carrier frequency of $\omega_0=2.0$~eV. The energy spectrum of this broadband pulse is depicted in the lower panel of Fig.~\ref{fig:fenBBpulse} by the red solid curve (velocity gauge). In the calculations, a random distribution of the CEP of this few-cycle pulse was assumed. Note that for chosen propagation times, the slow photoelectrons with kinetic energies below 0.4~eV cannot be separated from the highly-excited bound electronic states. \emph{Lower panel:} The respective multiphoton PECD in percent of the maximal intensity in the spectrum computed for one circular polarization (see caption of Fig.~\ref{fig:fenMulti} for details on the data presentation).}\label{fig:fenBBpecd}
\end{figure}

In order to simulate a random CEP of the broadband pulses, one has to perform calculations for different values of $\phi_0\in [0,2\pi)$ and average the spectra with equal weights. This makes the computational problem at hand almost unapproachable. Fortunately, in the case of equal appearance of all CEPs, such averaging can be performed analytically. In the very general case, a three-dimensional photoelectron momentum distribution can be presented via the full expansion over the spherical functions: $I(k,\theta,\varphi)=\sum_{LM} B_{LM}(k)\,Y_{LM}(\theta,\varphi)$. The origin of the azimuthal emission angle $\varphi$ can be set to the maximum of the electric field vector, which  is, in turn, defined by $\phi_0$. Therefore, integration of $I(k,\theta,\varphi)$ over the phase $\phi_0$ is equivalent to its integration over the angle $\varphi$. Only terms with $M=0$ in the full expansion survive such integration. In the other words, assuming a random distribution of CEP recovers the axial symmetry of the spectrum. It is therefore sufficient to average the momentum distributions, computed for one particular  $\phi_0$, over three orientation angles ($\alpha,\beta,\gamma)$ and to take its axially symmetric part $I(k,\theta)=\sum_{L} B_{L0}(k)\,Y_{L0}(\theta)$.

In the present calculations, we took $\phi_0=0$, set the maximum of the pulse envelope (\ref{eq:timesp}) at 7.5~fs, and propagated photoelectron wave packets over the full time of 27.5~fs. The somewhat longer propagation in the second half of the pulse was required to enable a better separation of photoelectrons with low kinetic energies from the highly-excited bound electronic states of the molecule. The peak intensity of the pulse was set to $3\times10^{12}$~W/cm$^2$. In order to support photoelectrons with the kinetic energies of up to about 15~eV, we used a radial grid of $r \leq 800$~a.u. Results of the present calculations are collected in Fig.~\ref{fig:fenBBpecd}.

The upper panel of Fig.~\ref{fig:fenBBpecd}  depicts the total energy spectrum of photoelectrons, released by the broadband pulses. One can assume that this spectrum is built of three decreasing in intensity and strongly-overlapping peaks, which are centered at about 1.1, 2.8, and 4.6 eV (note the two maxima and the shoulder at the respective kinetic energies). The almost constant energy separation of these three features by about 1.7~eV indicates that ionization proceeds in the intermediate regime between the narrow- and broadband extremes. At higher kinetic energies, the  spectrum decreases rapidly: It drops by two orders of magnitude at 8~eV (not shown in the figure). A reliable reconstruction of the low-energy part of the spectrum below 0.4~eV requires longer propagation times.

The lower panel of Fig.~\ref{fig:fenBBpecd}  depicts multiphoton PECD of fenchone obtained for the present coherent broadband pulses. It is shown in percent of the maximum in the spectrum (located at $\varepsilon \approx 1.0$~eV), computed for one of the polarizations. As one can see,  the utilized broadband pulses induce a substantial chiral asymmetry in the broad kinetic energy interval below 6~eV, and it is on the order of 10\% of the maximal ionization signal. The PECD exhibits an angular structuring, which is a fingerprint of the interference between different multiphoton ionization pathways. Interestingly, the computed multiphoton PECD changes its sign around the maximum of the first peak (at about 1.1~eV for the emission angle $\theta=0^\circ$ and around 1.9~eV for $\theta=90^\circ$).

\subsection{Bichromatic fields}
\label{sec:bichromatic}

In the recent theoretical work Ref.~\cite{PRLw2w}, it was  demonstrated that a substantial PECD can be induced by Lissajous-type bichromatic fields, which are linearly polarized in two mutually-orthogonal directions. The proposed field configuration of the $\omega$ and $2\omega$   pulses, which propagate along the laboratory $z$-axis, reads:
\begin{equation}
\mathcal{\vec{E}}(t)= \hat{e}_x \mathcal{E}_x g(t) \cos(2\omega t) + \hat{e}_y \mathcal{E}_y g(t)  \cos(\omega t +\phi).
\label{eq:fieldBI}
\end{equation}
For the relative phase $\phi=\pm \frac{\pi}{4}$ between the two fields, the electric field (\ref{eq:fieldBI}) follows the shape of a `butterfly' that is oriented along the $\omega$-field (see Fig.~1 in Ref.~\cite{PRLw2w}). Here, the electric field vector mimics two different rotational directions in the upper and lower parts of the dipole $xy$-plane, inducing  thereby chiral asymmetries of the opposite signs.  Reference~\cite{PRLw2w} demonstrated a sizable PECD which can be controlled by the relative phase $\phi$ between two fields.

In Ref.~\cite{PRLw2w}, calculations were performed for the two-red-photon vs one-blue-photon ionization of a model methane-like chiral system \cite{TDSC1} with high-frequency UV/XUV pulses. In addition, a possibility to utilize a higher-order multiphoton absorbtion regime was demonstrated on the example of six-red-photon vs three-blue-photon ionization of the same model system. Very recently \cite{Rozen19}, experiments proposed in Ref.~\cite{PRLw2w} were realized by studying the eight-red-photon vs four-blue-photon ionization of fenchone and camphor with  1030 and 515~nm IR/optical pulses. Those experiments were supported by model calculations of the six-red-photon vs three-blue-photon ionization of a toy chiral system with 800 and 400~nm optical pulses.

In this section, we report first calculations of the multiphoton PECD with bichromatic fields for real chiral molecule fenchone induced by optical laser pulses. In particular, we study six-red-photon vs three-blue-photon ionization of fenchone by the  800 and 400~nm optical pulses with $\phi=+\frac{\pi}{4}$ (`butterfly' field configuration), which is expected to induce a maximal chiral asymmetry. Because of the essential absence of the axial symmetry of such `butterfly' fields, electron spectra computed at different molecular orientations need to be averaged again over all three orientation Euler angles ($\alpha,\beta,\gamma)$. In order to reduce computational efforts, we covered orientation intervals of $\alpha,\gamma\in[0,2\pi)$ and $\beta\in[0,\pi]$ in steps of $\Delta \alpha= \Delta \beta=\Delta \gamma =0.2\,\pi$ and averaged computational results only over 500 molecular orientations.

To further decrease computational efforts, we used short sine-squared pulses with the full duration of about 22~fs. Such pulses support exactly 8 and 16 optical cycles for the wavelengths 800 and 400~nm, respectively. In order to avoid undesired contributions to the computed PECD signal from CEP asymmetry effects, we ensured  that the total ionization yields in the upper and lower hemispheres were equal: $I(k_y>0)=I(k_y<0)$. This was realized by setting a global phase such that both fields vanish at the pulse envelope maximum. In the calculations, we concentrated on the main PI signal in the electron spectrum, which is  centered at the kinetic energy of about $\varepsilon=0.7$~eV. Therefore, we used a small  radial grid with $r \leq 325$~a.u., which supports electrons with moderate kinetic energies, including 4ATI of the 400~nm pulse (8ATI of the 800~nm). In order to avoid the Autler-Towns splitting in the electron spectrum which is caused by the 800~nm pulse (see discussion in Sec.~\ref{sec:multiphoton}), we considerably reduced its peak intensity down to $10^{11}$~W/cm$^2$. Finally, to make contributions of both pulses to the threshold peak in the electron spectrum comparable, the peak intensity of the 400~nm pulse was set to $4\times 10^{11}$~W/cm$^2$.

\begin{figure}
\includegraphics[scale=0.6]{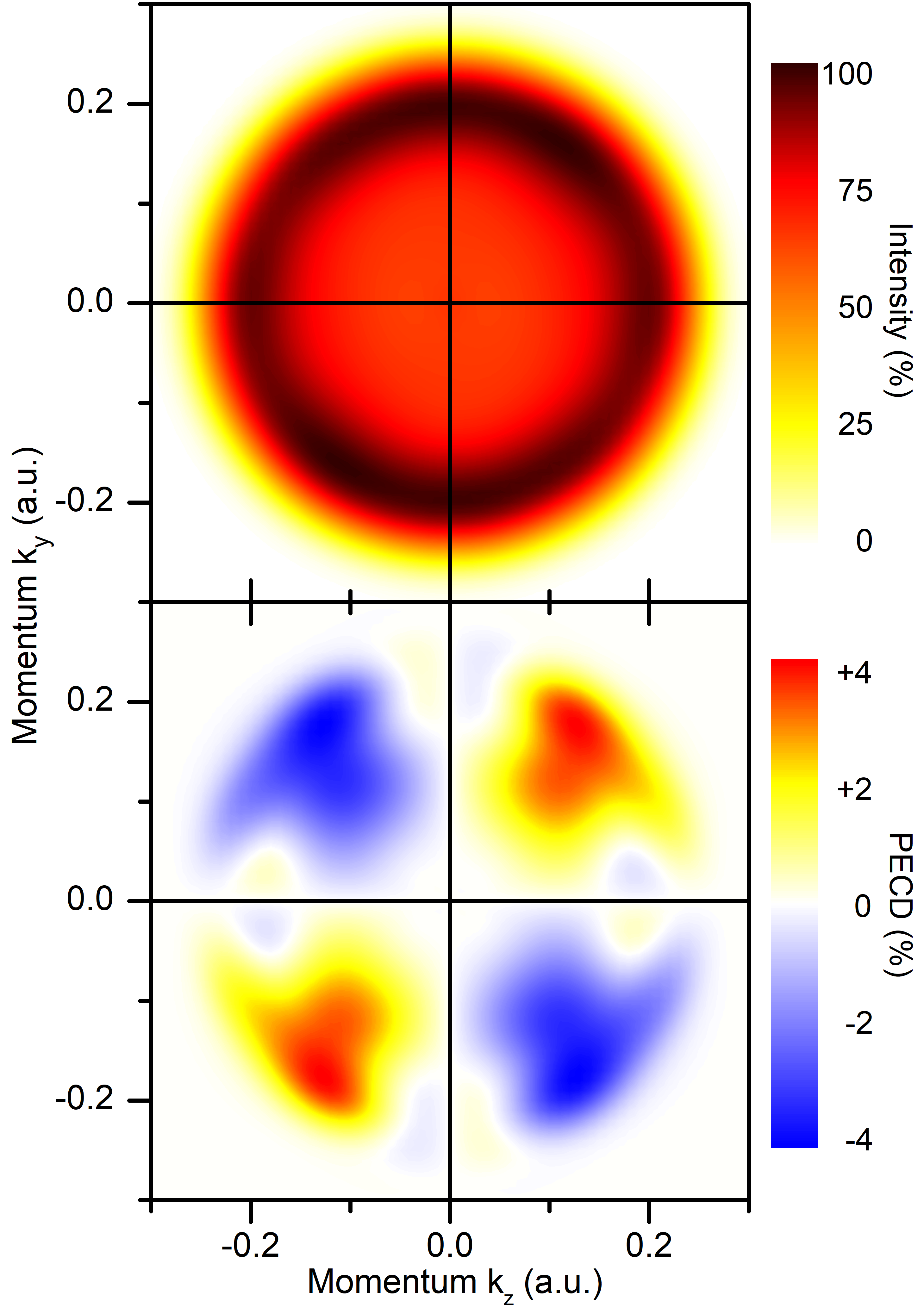}
\caption{Projection of the photoelectron angular distribution on the detector $yz$--plane (upper panel) and the respective PECD (lower panel) computed  in the present work for the six-red-photon vs three-blue-photon ionization (the first peak in the electron spectrum) of randomly-oriented R(--) fenchone molecules by the 800 and 400~nm bichromatic laser pulses Eq.~(\ref{eq:fieldBI}) and the relative phase $\phi=+\frac{\pi}{4}$ (butterfly form). The pulses propagate horizontally along the $k_z$-axis; the 800~nm field is linearly polarized along the $k_y$-axis; the projections are obtained by integrating the three-dimensional momentum distribution along the $k_x$-direction, which is the polarization direction of the 400~nm field. The PECD is obtained as the difference between the two signals $I(k_y,k_z)-I(-k_y,k_z)$ and shown in percent relatively to the maximal pixel intensity in the spectrum (set to 100\%). Note that  data are shown here on the linear scale in the photoelectron momentum $k$.}\label{fig:fenbichr}
\end{figure}

Results of the present calculations are collected in Fig.~\ref{fig:fenbichr}.  The upper panel of this figure depicts the projection of the three-dimensional photoelectron angular distribution on a detector placed in the $yz$-plane (i.e., perpendicular to the $2\omega$-field; see also Fig.~1 in Ref.~\cite{PRLw2w}). Note that only contribution from the strongest threshold peak produced by superposition of the 6PI(800~nm) and 3PI(400~nm) processes is depicted. This projection exhibits a noticeable asymmetry  for positive/negative values of $k_z$ (forward/backward directions with respect to the pulse propagation direction). The asymmetry is different for the positive/negative values of $k_y$ (upper/lower hemispheres), where for $\phi=+\frac{\pi}{4}$, the combined electric field vector (\ref{eq:fieldBI}) rotates in the opposite directions.

The lower panel of Fig.~\ref{fig:fenbichr} depicts PECD in percent of the maximal pixel intensity of the spectrum. It is obtained by subtracting the spectrum observed in the lower hemisphere from that in the upper one, and vise versa: $I(k_y,k_z)-I(-k_y,k_z)$. As one can see, the computed chiral asymmetry in the lower panel of Fig.~\ref{fig:fenbichr} reaches about 4\% of the maximal PI signal and possesses opposite signs in the upper/lower hemispheres. The PECD is also angularly structured, which is a clear indication of the higher-order (6PI vs 3PI) multiphoton pathways involved in the ionization process. A somewhat weaker chiral asymmetry well below 1\% was observed in Ref.~\cite{Rozen19} for the { large intensity multiphoton and strong-field} eight-red-photon vs four-blue-photon ionization of fenchone and camphor with the 1030 and 515~nm IR/optical pulses.

In the very recent theoretical work \cite{Demekhin19bi}, it was analytically demonstrated that in addition to the external phase $\phi$  between the two fields a molecule introduces an internal phase, which depends on the respective transition amplitudes for the six- and three-photon ionization. As a consequence, the bichromatic multiphoton PECD  maximizes at different external phases rather than $\phi=\pm\frac{\pi}{4}$. An influence of the internal molecular phase is especially pronounced for the resonance-enhanced multiphoton ionization schemes, as is likely the case here. Therefore, it can be expected that the multiphoton PECD, computed here for the `butterfly' field configuration, does not reach its maximal value. Searching for the external relative phase $\phi$, which maximizes PECD of fenchone in the considered process, is a very time-consuming computational task, and it is outside the scope of the present work.

\section{Conclusions}
\label{sec:Summary}

The time-dependent single center method, developed in Paper~I~\cite{TDSC1} and tested on bicyclic ketones in Paper~II~\cite{TDSC2}, is applied here to study the multiphoton PECD of fenchone in different  ionization regimes. The present multiphoton PECDs of the 3PI, 4ATI, and 5ATI electrons, computed for the 400~nm pulses, are in a very good agreement with the experimental results from Refs.~\cite{Lux12AngChm,Lux15CPC,Lux16ATI,Beaulieu16NJP,Beaulieu16td}. For the 814~nm pulses, the agreement of the present theory and the experiments of Refs.~\cite{Beaulieu16NJP,Beaulieu16td}, which utilize 800~nm pulses, is rather good. In particular, the present theory reproduces sizes and signs of the multiphoton PECDs of the 7ATI, 8ATI, and 9ATI electrons, while for the weakest 10ATI signal, the computed and measured PECDs have opposite signs. For strong-field ionization by the 1850~nm pulses, agreement of the present calculations with the measurements of  Refs.~\cite{Beaulieu16NJP,Beaulieu16td} is satisfactory. In particular, the calculations yield a weak multiphoton PECD of about 2\%, which is similar to the maximal asymmetry observed in Refs.~\cite{Beaulieu16NJP,Beaulieu16td}. However, the measured chiral asymmetry has one sign for all photoelectron kinetic energies, while the computed has not. {The present calculations support the theoretical conclusion of Ref.~\cite{Beaulieu16NJP} that, unlike a commonly accepted picture of tunnel ionization, PECD in this regime is caused by the effect of the final state scattering on the chiral potential of a molecule.}

The present computational results, obtained for the broadband ionization and multiphoton ionization by  bichromatic `butterfly' laser pulses, can be considered as theoretical predictions. For the 413--1240~nm spanning coherent pulses, we observe a sizable PECD in the large kinetic energy interval, and the computed chiral asymmetry reaches about 10\% of the maximal intensity of the spectrum. The PECD, computed for the six-red-photon vs three-blue-photon ionization by the 800 and 400~nm bichromatic pulses linearly polarized in two mutually-orthogonal directions, reaches about 4\%. The present theoretical results extend the \emph{universality} of the multiphoton PECD, proposed in Refs.~\cite{Beaulieu16NJP} for one-color multiphoton ionization and proved in Refs.~\cite{Beaulieu16td,Comby16td,Beaulieu17as,Beaulieu18PXCD} for different two-color  pump-probe schemes, to any kind of pulses  whose rotationally-tailored electric fields possess circulation \cite{PRLw2w,Rozen19}.

\begin{acknowledgements}
T. Baumert, A. Senftleben, H. Lee, H. Braun, A. Kastner and T. Ring are gratefully acknowledged for many valuable discussions. This work was funded by the Deutsche Forschungsgemeinschaft (DFG) -- Projektnummer 328961117 -- SFB 1319 ELCH (Extreme light for sensing and driving molecular chirality, subproject C1).  Part of the calculations has been performed at the Lichtenberg-Hochleistungsrechner of the Technische Universit\"{a}t Darmstadt (Projects 0628 and 0660).
\end{acknowledgements}


\begin{thebibliography}{99}

\bibitem{Lux12AngChm}
C. Lux, M. Wollenhaupt, T. Bolze, Q. Liang, J. K\"{o}hler, C. Sarpe, and T. Baumert,  Angew. Chem. Int. Ed. \textbf{51}, 5001 (2012).

\bibitem{Lehmann13jcp}
C.S. Lehmann, N.B. Ram, I. Powis, and M.H.M. Janssen, J. Chem. Phys. \textbf{139}, 234307 (2013).

\bibitem{Ram13EPJ}
N.B. Ram, C. S. Lehmann, and M.H.M. Janssen, EPJ Web Conf. \textbf{41}, 02029 (2013).

\bibitem{Lux15CPC}
C. Lux, M. Wollenhaupt, C. Sarpe, and T. Baumert,  Chem. Phys. Chem. \textbf{16}, 115 (2015).

\bibitem{Janssen14}
M.H.M. Janssen and I. Powis, Phys. Chem. Chem. Phys. \textbf{16}, 856 (2014).

\bibitem{Nanosec19}
A. Kastner, T. Ring, H. Braun, A. Senftleben and T. Baumert, ChemPhysChem \textbf{20}, 1416 (2019).

\bibitem{Lux16ATI}
C. Lux, A. Senftleben, C. Sarpe, M. Wollenhaupt, and T. Baumert, J. Phys. B \textbf{49}, 02LT01 (2016).

\bibitem{Beaulieu16NJP}
S. Beaulieu, A. Ferr\'{e}, R. G\'{e}neaux, R. Canonge, D. Descamps, B. Fabre, N. Fedorov, F. L\'{e}gar\'{e}, S. Petit, T. Ruchon, V. Blanchet, Y. Mairesse, and B. Pons,  New J. Phys. \textbf{18}, 102002 (2016).

\bibitem{Beaulieu16td}
S. Beaulieu, S. Comby, B. Fabre, D. Descamps, A. Ferr\'{e}, G. Garcia, R. G\'{e}neaux, F. L\'{e}gar\'{e}, L. Nahon, S. Petit, T. Ruchon, B. Pons, V.  Blanchet, and Y. Mairesse, Faraday Discuss. \textbf{194}, 325 (2016).

\bibitem{Cireasa15}
R. Cireasa, A. E. Boguslavskiy, B. Pons, M.C.H. Wong, D. Descamps, S. Petit, H. Ruf, N. Thir\'{e}, A. Ferr\'{e}, J. Suarez, J. Higuet, B.E. Schmidt, A. F. Alharbi, F. L\'{e}gar\'{e}, V. Blanchet, B. Fabre, S. Patchkovskii, O. Smirnova, Y. Mairesse, and V. R. Bhardwaj, Nat. Phys. \textbf{11}, 654  (2015).

\bibitem{Comby16td}	
A. Comby, S. Beaulieu, M. Boggio-Pasqua, D. Descamps, F. L\'{e}gare\'{e}, L. Nahon, S. Petit, B. Pons, B. Fabre, Y. Mairesse, and V. Blanchet, J. Phys. Chem. Lett. \textbf{7}, 4514 (2016).

\bibitem{Beaulieu17as}
S. Beaulieu, A. Comby, A. Clergerie, J. Caillat, D. Descamps, N. Dudovich, B. Fabre, R. G\'{e}neaux, F. L\'{e}gar\'{e}, S. Petit, B. Pons, G. Porat, T. Ruchon, R. Taïeb, V. Blanchet, and Y. Mairesse, Science \textbf{358}, 1288 (2017).

\bibitem{Beaulieu18PXCD}
{S. Beaulieu, A. Comby, D. Descamps, B. Fabre, G.A. Garcia, R. G\'{e}neaux, A.G. Harvey, F. L\'{e}gar\'{e}, Z. Ma\v{s}\'{\i}n, L. Nahon, A.F. Ordonez, S. Petit, B. Pons, Y. Mairesse, O. Smirnova, and V. Blanchet, Nat. Phys. \textbf{14}, 484 (2018).}

\bibitem{Rafiee16wl}	
M.M. Rafiee Fanood, M.H.M. Janssen, and I. Powis, J. Chem. Phys. \textbf{145}, 124320 (2016).

\bibitem{Kastner17wl}	
A. Kastner, T. Ring, B.C. Kr\"{u}ger, G.B. Park, T. Sch\"{a}fer, A. Senftleben,  and T. Baumert, J. Chem. Phys. \textbf{147}, 013926 (2017).

\bibitem{Comby16elee}
A. Comby, E. Bloch, C.M.M. Bond, D. Descamps, J. Miles, S. Petit, S. Rozen, J.B. Greenwood, V. Blanchet, and Y. Mairesse,
Nat. Commun. \textbf{9}, 5212 (2018).

\bibitem{Kastner16ee}
A. Kastner, C. Lux, T. Ring, S. Z\"{u}llighoven, C. Sarpe, A. Senftleben, and T. Baumert, ChemPhysChem \textbf{17}, 1119 (2016).

\bibitem{Dreissigacker14}
I. Dreissigacker and M. Lein, Phys. Rev. A. \textbf{89},  053406 (2014).

\bibitem{Goetz17}
R.E. Goetz, T.A. Isaev, B. Nikoobakht, R. Berger, and C.P. Koch, J. Chem. Phys. \textbf{146}, 024306 (2017).

\bibitem{Goetz19}
R.E. Goetz, C.P. Koch, and L. Greenman, Phys. Rev. Lett. \textbf{122}, 013204 (2019).

\bibitem{Rozen19}
S. Rozen, A. Comby, S. Beauvarlet, E. Bloch, D. Descamps, B. Fabre, S. Petit, V. Blanchet, B. Pons, N. Dudovich, and Y. Mairesse, Phys. Rev. X \textbf{9}, 031004 (2019).

\bibitem{TDSC1}
A.N. Artemyev, A.D. M\"{u}ller, D. Hochstuhl, and Ph.V. Demekhin, J. Chem. Phys. \textbf{142}, 244105 (2015).

\bibitem{TDSC2}
A.D. M\"{u}ller, A.N. Artemyev, and Ph.V. Demekhin, J. Chem. Phys. \textbf{148}, 214307 (2018).

\bibitem{Demekhin13H}
Ph.V. Demekhin, D. Hochstuhl, and L.S. Cederbaum, Phys. Rev. A \textbf{88}, 023422 (2013).

\bibitem{Artemyev16He1}
A.N. Artemyev, A.D. M\"{u}ller, D. Hochstuhl, L.S. Cederbaum, and Ph.V. Demekhin, Phys. Rev. A \textbf{93}, 043418 (2016).

\bibitem{Artemyev17He2}
A.N. Artemyev, L.S. Cederbaum, and Ph.V. Demekhin, Phys. Rev. A \textbf{95}, 033402 (2017).

\bibitem{Artemyev17He3}
A.N. Artemyev, L.S. Cederbaum, and Ph.V. Demekhin, Phys. Rev. A \textbf{96}, 033410 (2017).

\bibitem{Artemyev18He4}
A.N. Artemyev, A.I. Streltsov, and Ph.V. Demekhin,  Phys. Rev. Lett. \textbf{122}, 183201 (2019).

\bibitem{PRLw2w}
Ph.V. Demekhin, A.N. Artemyev, A. Kastner, and T. Baumert, Pys. Rev. Lett. \textbf{121}, 253201 (2018).

\bibitem{VG1}
E. Cormier and P. Lambropoulos, J. Phys. B \textbf{29}, 1667 (1996).

\bibitem{VG2}
Y.-C. Han and L. B. Madsen, Phys. Rev. A \textbf{81}, 063430 (2010).

\bibitem{PropL}
{T.J. Park and J.C. Light, J. Chem. Phys. \textbf{85}, 5870 (1986).}

\bibitem{Cont1}
Ph.V. Demekhin, D.V. Omelyanenko, B.M. Lagutin, V.L. Sukhorukov, L. Werner, A. Ehresmann, K.-H. Schartner, and H. Schmoranzer, Opt. Spectrosc. \textbf{102}, 318 (2007).

\bibitem{Cont2}
L.B. Madsen, L.A.A. Nikolopoulos, T.K. Kjeldsen, and J. Fern\'{a}ndez, Phys. Rev. A \textbf{76}, 063407 (2007).

\bibitem{Cont3}
Ph.V. Demekhin, A. Ehresmann, and V. L. Sukhorukov, J. Chem. Phys. \textbf{134}, 024113 (2011).

\bibitem{Cont4}
S.A. Galitskiy, A.N. Artemyev, K. J\"{a}nk\"{a}l\"{a}, B.M. Lagutin, and Ph.V. Demekhin, J. Chem. Phys. \textbf{142}, 034306 (2015).

\bibitem{Cont5}
S. Gozem, A.O. Gunina, T. Ichino, D.L. Osborn, J.F. Stanton, and A.I. Krylov, J. Phys. Chem. Lett. \textbf{6}, 4532 (2015).

\bibitem{Powis08}
I. Powis, C.J. Harding, G.A. Garcia, and L. Nahon, ChemPhysChem, \textbf{9} 475 (2008).

\bibitem{CoCo1}
M. Shapiro and P. Brumer, \emph{Quantum Control of Molecular Processes} (WILEY-VCH, Berlin, 2011).

\bibitem{CoCo2}
M. Wollenhaupt and T. Baumert, Faraday Discuss. \textbf{153}, 9 (2011).

\bibitem{Demekhin19bi}
Ph.V. Demekhin, Phys. Rev. A \textbf{99}, 063406 (2019).

\end{thebibliography}
\end{document}